\documentclass{elsart}

\input epsf.sty
\usepackage{amsmath}
\usepackage{amssymb}
\usepackage{graphics}
\usepackage{epsfig}

\begin{document}

\begin{frontmatter}



\title{Comments on the nuclear symmetry energy.}


\author{
Wojciech Satu{\l}a$^{1,2}$ and Ramon A. Wyss$^{1}$}

\address{$^1$Royal Institute of Technology, AlbaNova University Centre, 
106 91 Stockholm, Sweden}
\address{$^2$Institute of Theoretical Physics, University of Warsaw,
             ul. Ho\.za 69, PL-00-681 Warsaw, Poland}

\date{\today}

\maketitle

\begin{abstract}
 According to standard textbooks, the nuclear symmetry energy  originates
 from the {\it kinetic\/} energy and the {\it interaction\/} itself.
 We argue that this view requires certain modifications.
 We ascribe the physical origin of the {\it kinetic\/} term
 to the discreteness of fermionic levels of, in principle
 arbitrary binary fermionic systems, and relate its mean value
 directly to the average level density. Physically it connects this part 
 also to the isoscalar part of the interaction which, at least in 
 self-bound systems like atomic nuclei, decides upon the spatial 
 dimensions of the system. For the general case of binary fermionic systems
 possible external confining potentials as well as  specific boundary 
 conditions will contribute to this part. The reliability of this 
 concept is tested using self-consistent Skyrme Hartree-Fock calculations.
\end{abstract}

\begin{keyword}
Symmetry energy \sep Hartree-Fock calculations \sep level density
\PACS 21.10.Dr \sep 21.10.Pc \sep 21.60.Jz. 
\end{keyword}
\end{frontmatter}



  According to classic nuclear structure textbooks~\cite{[Boh69]}  
 the nuclear symmetry energy is composed of two basic ingredients: the
 so called {\it kinetic\/} and {\it interaction\/} parts:
\begin{equation}\label{esym}
       E_{sym} = \frac{1}{2} a_{sym} T^2 = 
                 \frac{1}{2} (a_{kin} +a_{int}) T^2.
\end{equation}
 This decomposition and terminology has its roots in the Fermi gas model.
 Within this model the strength of $a_{kin}$ can be easily evaluated 
 to be $a_{kin} \approx 100/A$\,MeV. The {\it interaction\/}
 part, on the other hand, is related to the Hartree part 
 of  a schematic isospin-isospin interaction: 
 \begin{equation}\label{tt} 
V_{TT}=
  \frac{1}{2}\kappa {\hat{\boldsymbol T}}\cdot {\hat{\boldsymbol T}}.
 \end{equation}
 The magnitude of $a_{int}$ (or alternatively interaction 
 strength $\kappa$) can be  estimated, however, only indirectly based on 
 the estimate of $a_{kin}$ and the actual knowledge of the
  empirical value of the total symmetry energy strength $a_{sym}$.

 The strength of the nuclear symmetry energy consists of a
 repulsive volume-type $\sim 1/A$ part which is reduced at
 the surface $\sim 1/A^{4/3}$. 
For the purpose of this work, 
 however, we adopt an effective  volume-like symmetry energy 
 strength $a_{sym}\approx 150/A$\,MeV. This strength must be 
 understood as the average over volume and surface contributions. 
 It is valid for relatively light nuclei
 of $A\sim 50\div 80$~\cite{[Mol95]}. Similar effective values 
 are obtained from fits assuming an $E_{sym}\sim T(T+1)$ dependence for the 
 total symmetry energy~\cite{[Duf95]}. Let us point out, however, 
 that none of the conclusions drawn below will depend upon the adopted 
 strength.



  An alternative way of calculating the symmetry energy is by means of
  the isospin cranking model~\cite{[Sat01a]}:
\begin{equation}\label{cran}
  \hat H^{\omega} = \hat H - {{\boldsymbol \omega}}
  \hat{{\boldsymbol T}}.
\end{equation}
  When the  
  nuclear hamiltonian ${\hat H}$ can be approximated by 
  an isospin symmetric single-particle ($sp$) mean-field 
  $h_{sp}=\sum_{occ} e_i$  the model can be strightforwardly solved 
  analytically. It  leads to the isospin parabolic energy law:
\begin{equation}\label{t2}
         E = \frac{1}{2} \varepsilon {T}^2
\end{equation}
  as shown in~Ref.~\cite{[Sat01a],[Sat01b],[Sat01c]}. 
  The formula (\ref{t2}) was formally derived  for an equidistant level 
  model characterized by a $sp$ splitting $\varepsilon$. Within 
  such a model the sequence of transitions from the $T=0$ ground state 
  to $T=2, 4,6,8,\cdots$ takes place at crossing frequencies which 
  form a simple arithmetic serie $\varepsilon, \, 
  3\varepsilon, \, 5\varepsilon, \cdots$. In reality, the 
  shell structure will affect that simple pattern leading to a sequence of
  nonuniformly spaced crossing frequencies determined by local 
  shell effects. Thus for small values of the isospin
 strong deviations from the 
  rule (\ref{t2}) may show up. However, for large values of $T$ the 
  uniform pattern  should be recovered with $\varepsilon$ being the
  measure of the  mean level spacing 
  at the Fermi energy provided that $\varepsilon (A,T_z)\approx 
  \varepsilon (A) $. 
Thus, from the perspective of the uniform isospin cranking model
the {\it 'kinetic'\/} symmetry-energy term originates rather from
the mean-level density than the {\it kinetic energy\/}.
Note that, apart from the smooth mean-part, this term also 
carries  information about the fluctuations, 
directly related to the local shell structure.

The mean-level spacing in nuclei can be rather easly evaluated. 
Including  Kramers and isospin degeneracy it is equal to 
$\varepsilon = 4/g(\lambda)$ 
where $g(\lambda)=6a/\pi^2$ stands for the mean density of states at the
Fermi energy. The empirical value of the level density parameter 
$a$, is uncertain. Typical estimates (again of
an effective volume-like $A$-dependent type) range between 
$a\approx A/10$\,MeV$^{-1}$ and $A/8$\,MeV$^{-1}$, 
see~\cite{[Gil65],[Kat80],[Shl92]}, where the lower limit for $a$ 
corresponds to the harmonic oscillator estimate~\cite{[Boh69]}.  
Thus the contribution is: 
\begin{equation}\label{mld}
 {\varepsilon} \approx
            {2\frac{\pi^2}{3a}} \approx {16\frac{\pi^2}{3A}} 
\div {20\frac{\pi^2}{3A}}  
 \approx  {\frac{53} {A}}\div {\frac{66}{A}}   \,\mbox{MeV}.
\end{equation}
Let us observe that using the Fermi gas model, the estimate for
$a=A/15$\,MeV$^{-1}$~\cite{[Boh69]} in Eq.~(\ref{mld})
recovers exactly the {\it kinetic\/} part of the symmetry energy 
strength $a_{kin}$ discussed in the introduction.
This value of the level density parameter is, however, 
unrealistic small.

  The repulsive isovector term in the nuclear potential (\ref{tt}) can be
  included within the mean-field isospin cranking model. Without loosing
  generality we can consider the one dimensional isospin cranking 
  model [e.g. around the $x$-axis, then $\langle \hat T_y\rangle =
   \langle \hat T_z\rangle =0$]. 
  After linearisation of the interaction (\ref{tt})  the  
  cranking Hamiltonian takes the following form:
\begin{equation}\label{mfham}
  \hat H_{MF}^{\omega} = \hat h_{sp} -
  ( \omega - \kappa \langle {\hat T_x}\rangle ){\hat T_x}
\end{equation}
  with an effective isospin dependent cranking frequency.  
  The isovector potential simply shifts the crossing frequencies 
  from $\varepsilon, 3\varepsilon, 5\varepsilon, \cdots$ to
   $\varepsilon +\kappa , 3(\varepsilon +\kappa), 
   5(\varepsilon+\kappa), \ldots$
  [within the equidistant level model] leading to  
\begin{equation}\label{t2_k}
         E = \frac{1}{2}(\varepsilon + 2\kappa) T^2.
\end{equation}
  The isovector field enters Eq.~(\ref{t2_k}) with  
  a factor of $2\kappa$, due to the restrictive 
  treatment of the linearized interaction as a {\it sp\/} external 
  potential $V_T = \kappa \langle \hat {\boldsymbol T}\rangle \hat
  {\boldsymbol T}$. 
  In this manner, the isovector 
 term is taken into account in phenomenological potentials.
Although $\kappa$ can be adjusted to yield the proper
Fermi energy dependence on $T$, summing up the total energy from the
occupied {\it sp\/} energies, results in a gross overestimate of the
symmetry energy.
  Hence, estimates of 
  $\kappa$ based on phenomenological potentials are not very
  reliable. 
Indeed, only by means of the Strutinsky shell-correction 
  procedure one is able to calculate the total energy. 
  This implies, however,  that the 
information carried out from the potential
  is stripped off any average trends  including the mean level 
  density and $\kappa$ in particular and only the fluctuations  
  are transfered.

  Nevertheless, the Strutinsky smooth energy can be used 
  for crude verification of the reliability and stability of 
  our concept. We calculated the smooth Strutinsky 
  energies for a chain of e-e $A$=88, $T_z=0\div 10$ nuclei. 
  We increased the strength of the 
isovector potential in steps from zero to its full value. 
  It turned out, that the level density, obtained in the calculations
does not depend on $T_z$.  This suggest that this method 
 can be viewed as an independent, accurate, and simple way of calculating 
the mean level density at the Fermi surface. 


%

  Within the Hartree-Fock (HF) approximation, the cranking
model (Eq.~(\ref{cran})) including an interaction $V_{TT}$ 
(Eq.~(\ref{tt})) will result in:
\begin{equation}\label{t2_hf}
         E = \frac{1}{2} (\varepsilon + \kappa) 
             { T}^2 +  \frac{1}{2}\kappa {T}.
\end{equation}

  Thus, the interaction (\ref{tt}) leads to a 
  linear term which comes entirely from the Fock exchange~\cite{[Nee02]} or,  
  in other words, results in   $E_{sym} \sim T(T+x)$ 
  with a $x< 1$ dependence for the symmetry energy. 
  This form of the symmetry energy is intermediate 
  between $E_{sym} \sim T^2$ and $E_{sym}\sim T(T+1)$ like
  the formulas debated in the literature. 
  Interestingly, the self-consistent HF or Hartree-Fock-Bogolyubov
  (HFB) models are believed to 
  give a quadratic $E_{sym}\sim T^2$ type dependence although, to the best of
  our knowledge, this conjucture has never been discussed seriously.  
  A dependence $\sim T(T+1)$, on the other hand, can be deduced from
  the nuclear shell-model.

  The interaction (\ref{tt}) is highly schematic as compared to 
  the isovector part of e.g. a Skyrme type 
  interaction.\footnote{To be precise, in the following, 
  the isovector part of the Skyrme 
  interaction always refers to that part of the interaction that 
  gives rise to the isovector dependence of the mean nuclear potential. 
  Within the standard Skyrme energy density functional it 
  is given by the isovector density dependent terms.} 
  Therefore, the
  symmetry energy formula (\ref{t2_hf}) can serve only as a  
  guideline for further investigations which are entirely based
  on the Skyrme-Hartree-Fock (SHF) energy functional. Nevertheless, 
  we will treat it literally and show that the entire  
  concept of strictly dividing the symmetry energy into 
  contributions from 
  {\it mean-level-density\/} and {\it interaction\/} 
  is  very reliable and highly transparent 
  also  from the point of self-consistent fields.
  To investigate the isovector effects of the strong force
  we switched off the Coulomb interaction and
  assumed the equality of proton and neutron masses.
  The calculations presented below were done using SHF code 
  HFODD of Ref.~\cite{[Dob00]} employing various parametrisations of 
  the Skyrme forces.

\begin{figure}
\centerline{\hbox{\psfig{file=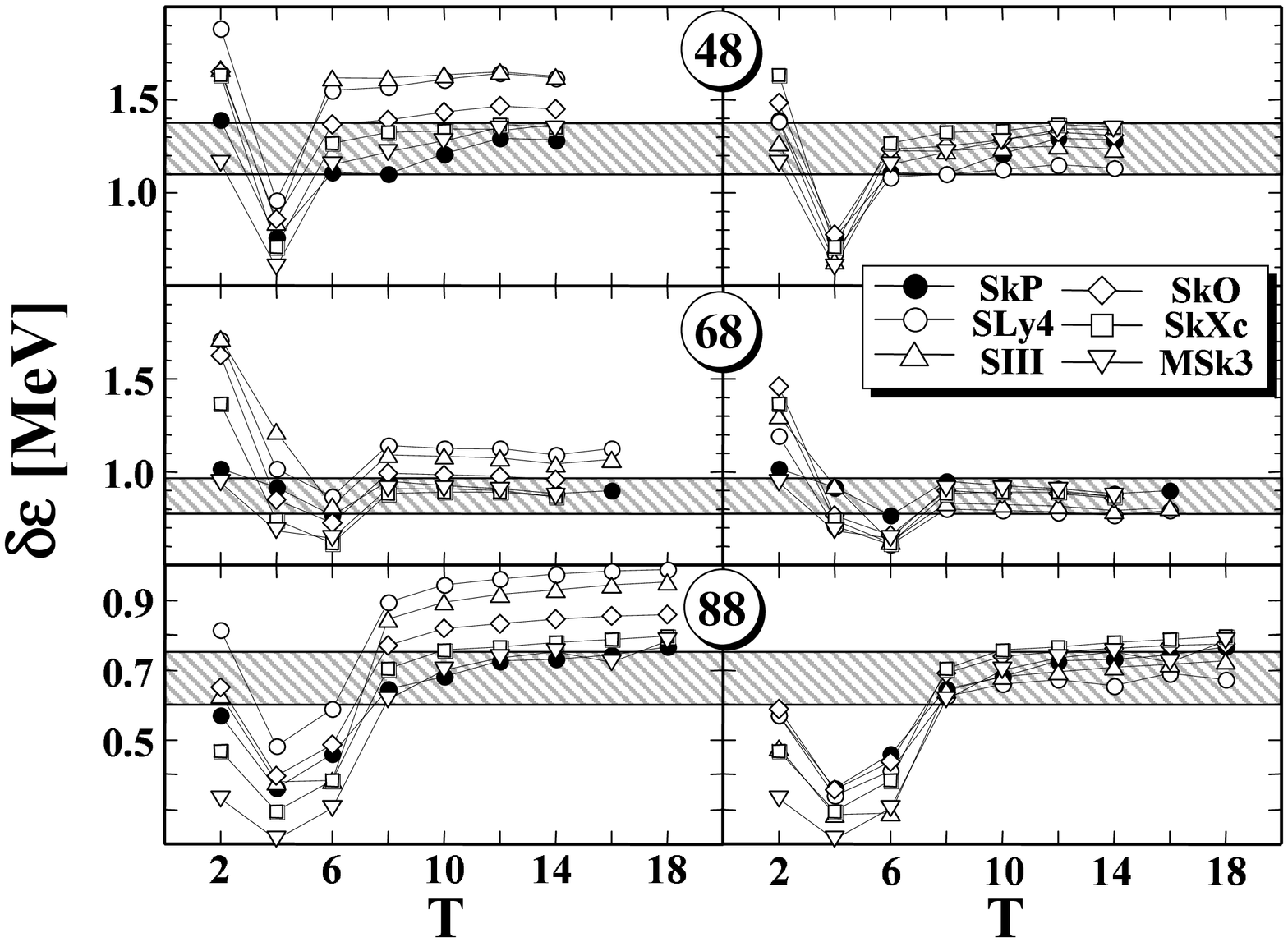,height=13.0cm,width=17.7cm,angle=0}}}
\mbox{}
\caption{
The mean-level spacing (left hand side) calculated using only the 
isoscalar part of the SHF interaction 
for the $A=48,68,88$ isobaric chains. Parametrisations used are 
indicated in the legend. Right hand side shows the results scaled 
by the isoscalar effective mass $\frac{m*}{m} \varepsilon$. Shaded 
areas indicate the estimates given by~\protect{(\ref{mld})}.}
\label{fig1}
\end{figure}

  The {\it mean-level-density\/} $\varepsilon$ contribution can 
  be studied simply by removing the isovector part of the Skyrme 
  force. Then, guided by Eq.~(\ref{t2_hf}), the calculated 
  HF energy is approximated by:
\begin{equation}
 \Delta E^{(HF)}_T = E^{(HF)}_{T} -  E^{(HF)}_{T=0} \approx
  \frac{1}{2}\varepsilon T^2,
\end{equation}
  which allows to extract $\varepsilon$ as a function of
  $T$($\equiv T_z$).
  The results of the calculations are presented in 
  Fig.~\ref{fig1}. The calculations have been done  
  for several Skyrme forces including SLy4~\cite{[sly]},
  SIII~\cite{[Bei75]} and SkO~\cite{[Rei99]} i.e. forces having 
  isoscalar effective masses equal $m^*/m \approx 0.70, 0.76$ 
  and 0.90, respectively, as well as 
   SkP~\cite{[Dob84]}, SkXc~\cite{[Bro98]} and MSk3~\cite{[Ton00]} 
  i.e. forces having $m^*/m \approx 1$.
  The calculations were done for the even-even $A=48, 68$ and 88 isobaric 
  chains of nuclei from $T_z=0$ to the vicinity of the drip line.

  As anticipated, for relatively small values of $T_z$ strong variations 
  in $\varepsilon$ are seen. However for larger values of $T_z\geq 8$
  the values of $\varepsilon$ stabilize. Let us point out that the forces 
  having effective masses considerably lower than unity have a
  substantially larger $\varepsilon$. However, after employing
  an effective mass scaling $\varepsilon \rightarrow 
  \frac{m^*}{m} \varepsilon$ all curves are within the limits 
  given by the estimate (\ref{mld}) [shaded areas in the figure].
  Clearly, the physical interpretation of this term as arising from the 
  level density is evident. 
  This observation nicely confirms the validity of our approach. 
  Let us further observe that, with increasing $A$, all curves move 
  towards the upper limit of (\ref{mld}).
  This, most likely, reflects the decreasing role of surface effects 
   with increasing $A$. Indeed, more accurate estimates of the average
  density of states $g$, including surface effects can be done based on
  the semiclassical theory~\cite{[Bal70]}. 
  It can be shown that surface effects 
  increase the density of states $g_{V+S} > g_{V}$ and, in turn,
  lower average level spacing $\varepsilon_{V+S} < \varepsilon_V$ thus
  giving rise to the observed trend. A detailed 
  discussion of the $A$ dependence of the symmetry energy based on 
  these ideas will be published elsewhere.

\begin{figure}
\centerline{\hbox{\psfig{file=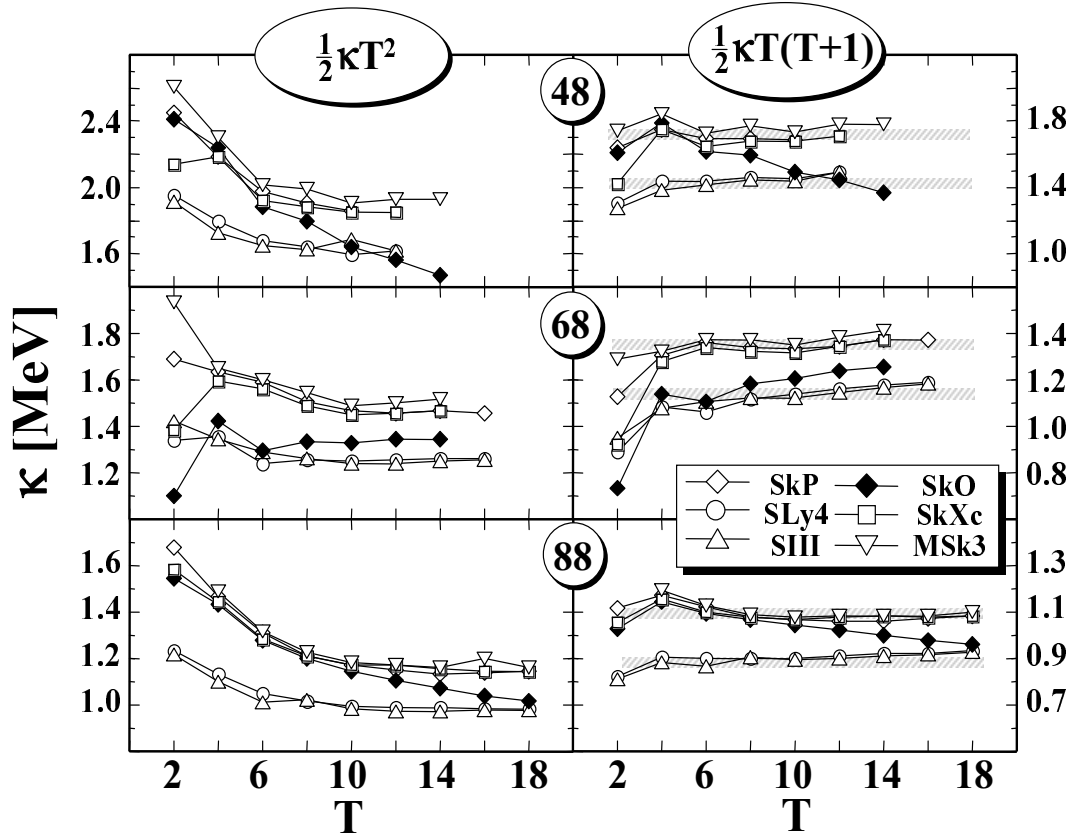,height=13.0cm,width=17.8cm,angle=0}}}
\mbox{}
\caption{
The average effective strength $\kappa$  of the isovector part of 
various Skyrme forces. Left hand side shows the values of
$\kappa$ estimated assuming $E_{sym} = (\varepsilon +\kappa)T^2/2$.
Right hand side shows the values of
$\kappa$ estimated assuming $E_{sym} = \varepsilon T^2/2 
+\kappa T(T+1)/2$. In both cases the values of $\varepsilon$ 
from the isoscalar SHF calculations have been subtracted.
}
\label{fig2}
\end{figure}

  The dependence of $\varepsilon$ on the effective mass 
  is directly related to the isovector part of the Skyrme interaction. 
  A small effective mass results in a reduction of the isovector Skyrme
  interaction, since the overall symmetry energy must reproduce, by
  construction, the empirical value. This effect is seen 
  in Fig.~\ref{fig2}  showing the estimated values of $\kappa$
  which, in this context, have the  meaning of an  
  {\it effective strength\/} of the isovector 
  component of the Skyrme force. We have performed two 
  set of calculation to extract $\kappa$.  In the calculations
  presented in the left panel of Fig.~\ref{fig2}, we have assumed 
  a purely quadratic isospin dependence 
  $\Delta E^{(HF)}_T = 
  \frac{1}{2}(\varepsilon +\kappa) T^2$. The right panel, on the 
  other hand,  shows 
  calculations that include also the linear term from
  Eq.~(\ref{t2_hf}). For both cases we have made use of the 
  $\varepsilon$ values 
  extracted previously (Fig.~\ref{fig1}) from the isoscalar SHF 
  calculations.The calculations reveal at least two striking features: 
  ({\it i\/}) No major 
  perturbations in the $\kappa (T_z)$ curves are seen for large $T_z$ 
  pointing to a very weak, if any, impact of the isovector fields on the 
  average level density. 
  ({\it ii\/}) Clearly, and rather surprisingly, these calculations 
  reveal the presence of a linear term. Indeed,
  the curves in the right panel of Fig.~\ref{fig2} show essentially no
  $T_z$  dependence. The only exception from this rule is the SkO 
  force with its highly non-standard strong, repulsive isovector
  component of the spin-orbit force.

  An interesting conjecture 
  can be made here concerning recent very accurate self-consistent 
  mass calculations based on Skyrme forces.~\cite{[Gor02]} 
  These calculations 
  clearly prefere Skyrme forces of $m^*/m\sim 1$. This preference 
  can be (partly) related to the presence of the linear term in 
  the symmetry energy, provided, that the empirical symmetry energy 
  has a term like $\sim T(T+x)$. 
  Indeed, for Skyrme forces having $m^*/m <1$ the linear term is expected  
  to be reduced.  On the other hand, there seems to be no 
  counterpart to the $\sim \frac{1}{2}\varepsilon T$ type at the 
  mean-field level. Such a compensating term may be %
  calculated within the RPA theory as suggested recently by~\cite{[Nee02]}.
  Performing RPA calculations atop of HF/HFB particularly in 
  large-scale mass calculations is presently beyond reach.  
  However, since the {\it linear\/} term describes, in fact, 
  the isospin dispersion, and since our interpretation links
  the inertia parameter to the average spacing
  $\varepsilon$, the overall correction 
  can easily be estimated based on Eq.~(\ref{mld}) 
  [scaled by the isoscalar effective mass] and the calculated isospin 
  dispersion $\langle \Delta \vec{\boldsymbol T}^2\rangle$.

%
%

\begin{figure}
\centerline{\hbox{\psfig{file=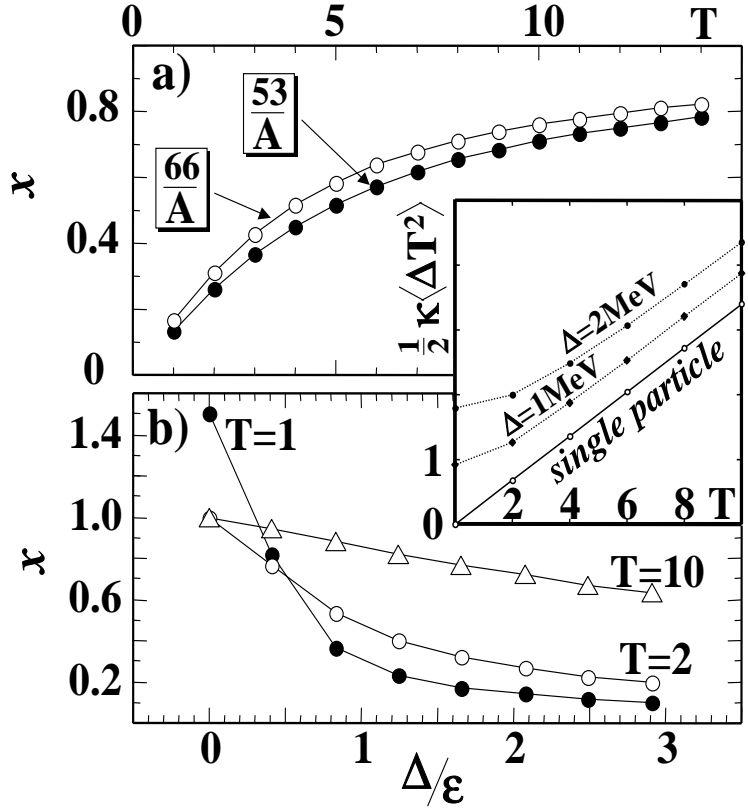,height=12.0cm,width=16.cm,angle=0}}}
\mbox{}
\caption{BCS calculations, based on a schematic, equidistant level model of the
parameter $x$ (Eq.~\protect{\ref{ex}})  in the {\it interaction\/} related
part of $E_{sym}^{int} \sim \kappa T(T+x)$. The upper part shows $x(T_z)$
calculated for $A=88$, $\Delta=12/\sqrt{A}$\,MeV and two values of
$\varepsilon$ as indicated in the figure. The lower part shows the
quenching of $x$ versus $\Delta/\varepsilon$ for $T_z=1$ [the filling
approximation is used], $T_z=2$ and $T_z=10$ nuclei.
The mechanism responsible for the quenching of $x$ in $N\sim Z$ nuclei
is displayed in the inset which shows the $T$-dependence
of the interaction related linear term $\kappa \langle
\Delta{\boldsymbol T^2}\rangle /2$ for $\Delta =$0 (single-particle) 1, and
2\,MeV, respectively. 
}
\label{fig3}
\end{figure}

In the
{\it sp\/} limit, one finds that 
$\langle \Delta  {\boldsymbol T}^2\rangle= T$ and therefore,
the {\it interaction\/} part of the symmetry energy
follows $E_{sym}^{int}\sim T(T+1)$. 
However, pairing correlations tend to increase
$\langle \Delta  {\boldsymbol T}^2\rangle$, with the largest
effect in $N$=$Z$ nuclei,
 as shown in the insert of Fig.~\ref{fig3}.
 Thus pairing effectively weakens the linear term leading to
$E_{sym}^{int} \sim T(T+x)$ with $x < 1$. The impact of
pairing on the value of $x$ is illustrated in Fig.~\ref{fig3},
and defined as
\begin{equation}\label{ex}
x \equiv \frac{\langle \Delta  {\boldsymbol T}^2\rangle_{T}
 - \langle \Delta  {\boldsymbol T}^2\rangle_{T=0}}{T}
 \quad \mbox{where} \quad T\equiv T_z = (N-Z)/2
\end{equation}
The value of $x$ is calculated  using the schematic, non-selfconsistent
constant gap $\Delta$ BCS calculations.
In the calculations the equidistant level spectrum characterized
by a {\it sp\/} splitting $\varepsilon$ was used. $2N$ ($2Z$) of the double
degenerate levels were active in the pairing calculations,
in order to take advantage of the particle-hole symmetry
which in turn guarantees automatic number conservation.

As already discussed pairing has the greatest impact on $N\sim Z$ nuclei
reducing the value of $x$ to $\sim 0.3$ for the $A=88, T_z=1$ system, see
Fig.~\ref{fig3}a. With inreasing $T$, the value of $x(T)$ increases slowly to
$x \rightarrow 1$. For small $T$ the actual value of $x$
changes rapidly with an increasing ratio of $\Delta/\varepsilon$
as illustrated in Fig.~\ref{fig3}b. Since
$\Delta/\varepsilon \sim A^\nu$ [$\nu \sim 1/2 \div 2/3$]
the quenching of $x$ is expected to be more efficient
in heavy $N\sim Z$ nuclei. Local shell-structure, however,
may result in oscillations of the $\Delta/\varepsilon$ ratio
and, in turn, to strong variations in $x$.

\begin{figure}
\centerline{\hbox{\psfig{file=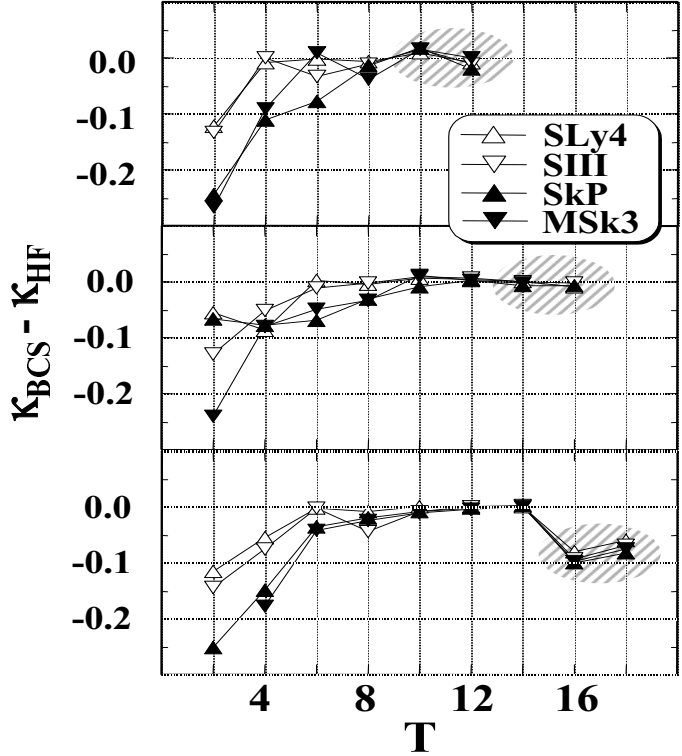,height=15.0cm,width=12.cm,angle=0}}}
\mbox{}
\caption{
The difference between HF+BCS and HF values of $\kappa$. In both
cases $\kappa$ has been calculated assuming  
$E_{sym}^{int} \sim \kappa T(T+1)$. Note that values of $\kappa_{BCS} -
\kappa_{HF}$ are always negative. This behavior shows
that pairing has a similar destructive impact on the {\it linear\/} 
Skyrme related term as emerged from the simple schematic 
considerations shown in Fig.~\protect{\ref{fig3}}. 
}
\label{fig4}
\end{figure}

To verify this assumption  we performed self-consistent HF+BCS 
calculations using SLy4, SIII, SkP and MSk3 forces.
The calculation scheme was similar to the one described 
above. First we extracted $\varepsilon_{BCS}$ from
the isoscalar Skyrme SHF+BCS calculations. Then,
using these values, we extracted $\kappa_{BCS}$.
The values of $\kappa_{BCS}$ were computed assuming 
$E_{sym}^{int} \sim \kappa T(T+1)$ and compared to 
corresponding values from pure SHF calculations,
$\kappa_{HF}$, see Fig.~\ref{fig2}. The difference 
is shown in Fig.~\ref{fig4}. It is always negative,
and particularly large  for small $T_z$. This 
clearly confirms the schematic
model calculations shown in Fig.~\ref{fig3}.

In conclusion, it is
demonstrated that the Skyrme forces 
give rise to a linear term in the {\it interaction\/}
part of $E_{sym}$ which behaves effectively 
similar to the schematic interaction (\ref{tt}), i.e. 
like $\sim \langle \Delta {\boldsymbol T}^2\rangle$.  
This term is strongly quenched in $N\sim Z$ nuclei
due to $pp/nn$ pairing, which, even after including RPA 
correlations~\cite{[Nee02]}, seem to leave 
room for
isoscalar pairing correlations in $N=Z$ nuclei. 
Particularly, since the data clearly indicates that 
the linear term of the symmetry energy in the vicinity of the $N=Z$ line is 
enhanced and  goes like $\sim T(T+1.25)$~\cite{[Jan65],[Glo02]}. 
Mass calculations with the SHF need to take into account the
fluctuations in isospin, $\sim \langle \Delta {\boldsymbol T}^2\rangle$
similar to the fluctuations in spin~\cite{[Ton00]} with an inertia
given by the mean level spacing at the Fermi energy. This may allow for more
general Skyrme forces, not restricted to effective 
mass $m^*\approx m$.

In particular, the present investigation reveals
that one component of the nuclear symmetry energy 
can directly be associated with	
the {\it mean-level density\/}
rather then the {\it kinetic\/} energy as proposed
in textbooks.  
Although the arguments are based on 
Skyrme-Hartree-Fock calculations of 
$A=48,68$, and 88 chains of nuclei, 
this is a general feature 
originating from 
the granularity of the fermionic single
particle spectra
and not
related to atomic nuclei only.

\bigskip

This work was supported by the G\"oran Gustafsson Foundation,
the Swedish Natural Science Research council (NFR),
and the Polish Committee for Scientific Research (KBN) under
Contract No. 5~P03B~014~21

\bibliographystyle{unsrt}

\end{document}